\documentclass[prb,twocolumn,showpacs,superscriptaddress,amsmath,amssymb]{revtex4}
\usepackage{graphicx}

\begin{document}
\title{Local electronic structure and magnetic properties of LaMn$_{0.5}$Co$_{0.5}$O$_3$
studied by x-ray absorption and magnetic circular dichroism
spectroscopy}

\author{T. Burnus}
\affiliation{II. Physikalisches Institut, Universit\"at zu K\"oln,
Z\"ulpicher Stra\ss e 77, 50937 K\"oln, Germany}
\author{Z. Hu}
\affiliation{II. Physikalisches Institut, Universit\"at zu K\"oln,
Z\"ulpicher Stra\ss e 77, 50937 K\"oln, Germany}
\email{zhiwei@ph2.uni-koeln.de}
\author{H. H. Hsieh}
\affiliation{Chung Cheng Institute of Technology, National Defense
University, Taoyuan 335, Taiwan}
\author{V. L. J. Joly}
\affiliation{Physical and Materials Chemistry Division, National
Chemical Laboratory, Pune 411008, India}
\author{P. A. Joy}
\affiliation{Physical and Materials Chemistry Division, National
Chemical Laboratory, Pune 411008, India}
\author{M. W. Haverkort}
\affiliation{II. Physikalisches Institut, Universit\"at zu K\"oln,
Z\"ulpicher Stra\ss e 77, 50937 K\"oln, Germany}
\author{Hua Wu}
\affiliation{II. Physikalisches Institut, Universit\"at zu K\"oln,
Z\"ulpicher Stra\ss e 77, 50937 K\"oln, Germany}
\author{A.~Tanaka}
\affiliation{Department of Quantum Matter, ADSM, Hiroshima
University, Higashi-Hiroshima 739-8530, Japan}
\author{H.-J. Lin}
\affiliation{National Synchrotron Radiation Research Center, 101
Hsin-Ann Road, Hsinchu 30076, Taiwan}
\author{C. T. Chen}
\affiliation{National Synchrotron Radiation Research Center, 101
Hsin-Ann Road, Hsinchu 30076, Taiwan}
\author{L. H. Tjeng}
\affiliation{II. Physikalisches Institut, Universit\"at zu K\"oln,
Z\"ulpicher Stra\ss e 77, 50937 K\"oln, Germany}

\date{\today}

\begin{abstract}
We have studied the local electronic structure of
LaMn$_{0.5}$Co$_{0.5}$O$_{3}$ using soft-x-ray absorption
spectroscopy at the Co-$L_{3,2}$ and Mn-$L_{3,2}$ edges. We found
a high-spin Co$^{2+}$--Mn$^{4+}$ valence state for samples with
the optimal Curie temperature. We discovered that samples with
lower Curie temperatures contain low-spin nonmagnetic Co$^{3+}$
ions. Using soft-x-ray magnetic circular dichroism we established
that the Co$^{2+}$ and Mn$^{4+}$ ions are ferromagnetically
aligned. We revealed also that the Co$^{2+}$ ions have a large
orbital moment: $m_{\rm orb}$/$m_{\rm spin}\approx0.47$. Together
with model calculations, this suggests the presence of a large
magnetocrystalline anisotropy in the material and predicts a
nontrivial temperature dependence for the magnetic
susceptibility.
\end{abstract}

\pacs{
71.27.+a, 
78.70.Dm, 
71.70.-d, 
75.25.+z  
}

\maketitle


The manganites continue to attract considerable attention from
the solid state physics and chemistry community over the last
five decades because of their spectacular material
properties.\cite{Jonker1950,Volger1954,Ramirez1997,Imada1998} The
parent compound LaMnO$_3$ is an A-type antiferromagnetic
insulator with orthorhombic perovskite crystal structure.
Replacing La by Sr, Ca or Ba results in multifarious electronic
and magnetic properties including the transformation into a
ferromagnetic state accompanied by a metal-insulator transition
and the occurrence of colossal
magnetoresistance.\cite{Helmolt1993,Jin1994} Substitution of the
magnetic Mn ions by Co also yields ferromagnetism in the
LaMn$_{1-x}$Co$_x$O$_3$ series. The Curie temperature reaches a
maximum for $x=0.5$
($T_C=220$--240~K).\cite{Goodenough1961,Blasse1965,Jonker1966,Herbert2002,Dass2003}
This should be contrasted with the end member of this series,
namely the rhombohedral LaCoO$_3$, which is a nonmagnetic
insulator at low temperatures, showing yet the well-known
spin-state transition at higher temperatures which by itself is
subject of five decades of intensive
study.\cite{Goodenough1961,Jonker1966,Haverkort2006}

Explaining the appearance of ferromagnetism in the manganites by
Co substitution is, however, not a trivial issue. Assuming that
ordering of the Co and Mn ions had not been achieved for the
$x=0.5$ composition, Goodenough \textit{et al.} concluded early on
that the ferromagnetism is generated by Mn$^{3+}$--O--Mn$^{3+}$
superexchange interactions.\cite{Goodenough1961} On the other
hand, later magnetic susceptibility and Mn NMR studies suggested
that it is the exchange interaction involving the ordering of
Co$^{2+}$--Mn$^{4+}$ transition-metal ions which causes the
ferromagnetism in
LaMn$_{0.5}$Co$_{0.5}$O$_{3}$.\cite{Blasse1965,Jonker1966,Nishimori1995,Troyanchuk1997,Troyanchuk1997b,Asai1998,Troyanchuk2000}

Only few high-energy spectroscopic studies are reported for the Co
substituted manganites. Using soft-x-ray absorption spectroscopy
(XAS), Park \textit{et al.} found in their low Co compositions
that the Co ions are divalent, favoring a Mn$^{3+}$--Mn$^{4+}$
double-exchange mechanism for the ferromagnetism.\cite{Park1997}
Extrapolating this Co divalent result to the $x=0.5$ composition
would provide support to the suggestion that the ferromagnetism
therein is caused by the Co$^{2+}$--Mn$^{4+}$ exchange
interaction. However, no XAS data have been reported so far for this
$x=0.5$ composition. Using $K$-edge XAS, Toulemonde \textit{et
al.} revealed that the Co ion is also divalent in their hole
doped and Co substituted manganite.\cite{Toulemonde1998} Yet,
these results for the low Co limit have been questioned by van
Elp, who claimed that the Co ions should be in the
intermediate-spin trivalent state rather than in the high-spin
divalent state.\cite{vanElp1999}

Further discussion is also raised by the work of Joy and
coworkers,\cite{Joy2000a,Joly2001b} who have synthesized two different
single phases of LaMn$_{0.5}$Co$_{0.5}$O$_{3}$ and inferred from
a combination of magnetic susceptibility and x-ray photoelectron
spectroscopy measurements that the phase with the higher $T_C$
contains high-spin Mn$^{3+}$ and low-spin Co$^{3+}$ ions, while
the lower $T_C$ phase has Co$^{2+}$ and Mn$^{4+}$. Very recently,
however, long-range charge ordering has been observed in neutron
diffraction experiments by Bull \textit{et al.}\cite{Bull2003} and Troyanchuk
\textit{et al.}\cite{Troyanchuk2004} on the high-$T_C$ phase, pointing towards the
Co$^{2+}$--Mn$^{4+}$ scenario. Also,
the most recent magnetic susceptibility and $K$-edge XAS data by
Ky\^omen \textit{et al.} favor the presence of essentially
Co$^{2+}$--Mn$^{4+}$ at low temperatures.\cite{Kyomen2003} The
issue of Mn/Co ordering including the possible coexistence of
ordered and disordered regions remains one of the important
topics.\cite{Dass2003,Kyomen2004,Autret2005} Interesting is that
the magnetization of polycrystalline samples of
LaCo$_{0.5}$Mn$_{0.5}$O$_3$ does not saturate in magnetic fields
up to 7 T,\cite{Asai1998} and that there are indications for a
large magnetic anisotropy.\cite{Troyanchuk2004}

On the theoretical side, not much work has been carried out so
far. A relatively early band-structure study by Yang \textit{et
al.} on the LaMn$_{0.5}$Co$_{0.5}$O$_{3}$ system predicted a
half-metallic behavior with a magnetic moment of $3.01\mu_B$ for
Mn and $0.54\mu_B$ for Co ions, suggesting Mn$^{3+}$--Co$^{3+}$
valence states.\cite{Yang1999} This study, however, was performed
before the existence of the charge-ordered crystal structure was
reported.\cite{Bull2003,Troyanchuk2004}

Here, we present our experimental study of the local electronic
structure of LaMn$_{0.5}$Co$_{0.5}$O$_{3}$ both for the high- and
low-$T_C$ phases using the element-specific XAS and x-ray magnetic
circular dichroism (XMCD) at the Co-$L_{2,3}$ and Mn-$L_{2,3}$
edges, i.e., transitions from the $2p$ core to the $3d$ valence
orbitals. Our objective is not only to establish the valence and
spin states of the Co and Mn ions but also to investigate the
possible presence of an orbital moment associated with a
Co$^{2+}$ ion, in which case the material should have a large
magnetocrystalline anisotropy and a nontrivial temperature
dependence of its magnetic susceptibility.

In XAS and XMCD we make use of the fact that the Coulomb
interaction of the $2p$ core hole with the $3d$ electrons is much
larger than the $3d$ band width, so that the absorption process
is strongly excitonic and therefore well understood in terms of
atomiclike transitions to multiplet-split final
states.\cite{Tanaka94,deGroot94,Thole97} Unique to soft-x-ray
absorption is that the dipole selection rules are very effective
in determining which of the $2p^{5}3d^{n+1}$ final states can be
reached and with what intensity, starting from a particular
$2p^{6}3d^{n}$ initial state ($n=7$ for Co$^{2+}$, $n=6$ for
Co$^{3+}$, $n=4$ for Mn$^{3+}$, and $n=3$ for Mn$^{4+}$). This
makes the technique an extremely sensitive local probe, ideal to
study the valence\cite{Chen90,Mitra2003} and
spin\cite{Laan88,Thole88,Cartier92,Pen97,Hu04,Haverkort2006}
character as well as the orbital contribution to the magnetic
moment\cite{Thole1992,Chen95,Burnus2006} of the ground or initial
state.


The two single-phase LaMn$_{0.5}$Co$_{0.5}$O$_{3}$ polycrystalline samples
were synthesized as described previously\cite{Joy2000,Joy2000a,Joly2001b}
and the single phase nature of the two phases (low-$T_C$ phase and high-$T_C$ phase) were confirmed by temperature dependent magnetization measurements. These measurements showed a single sharp magnetic transition at $T_C = 225$~K (called high-$T_C$ phase) for the sample synthesized at 700 $^\circ$C and a sharp transition at $T_C = 150$ K (called low-$T_C$ phase) for the sample synthesized at 1300 $^\circ$C. On the other hand, more than one magnetic transition or broad magnetic transitions were observed for samples synthesized at other temperatures indicating their mixed phase behavior, as described in Ref. \onlinecite{Joy2000}. The magnetization at 5 K in a field of 5 T is 50.4 emu/g for the high-$T_C$ phase and 42.4 emu/g for the low-$T_C$ phase.
The Co- and Mn-$L_{2,3}$ XAS and XMCD
spectra were recorded at the Dragon beamline of the National
Synchrotron Radiation Research Center (NSRRC) in Taiwan with an
energy resolution of 0.25~eV. The sharp peak at 777.8~eV of the
Co-$L_3$ edge of single crystalline CoO and at 640~eV of the
Mn-$L_3$ of single crystalline MnO were used for energy
calibration. The isotropic XAS spectra were measured at room
temperature, whereas the XMCD spectra at both the Co-$L_{2,3}$ and the
Mn-$L_{2,3}$ edges were measured at 135~K in a 1 T magnetic field
with approximately 80\% circularly polarized light. The magnetic
field makes an angle of $30^\circ$ with respect to the Poynting
vector of the soft x-rays. The spectra were recorded using the
total electron yield method (by measuring the sample drain
current) in a chamber with a base pressure of $2\times10^{-10}$ mbar.
Clean sample areas were obtained by cleaving the polycrystals
\textit{in situ}.

\begin{figure}
    \includegraphics[width=0.5\textwidth]{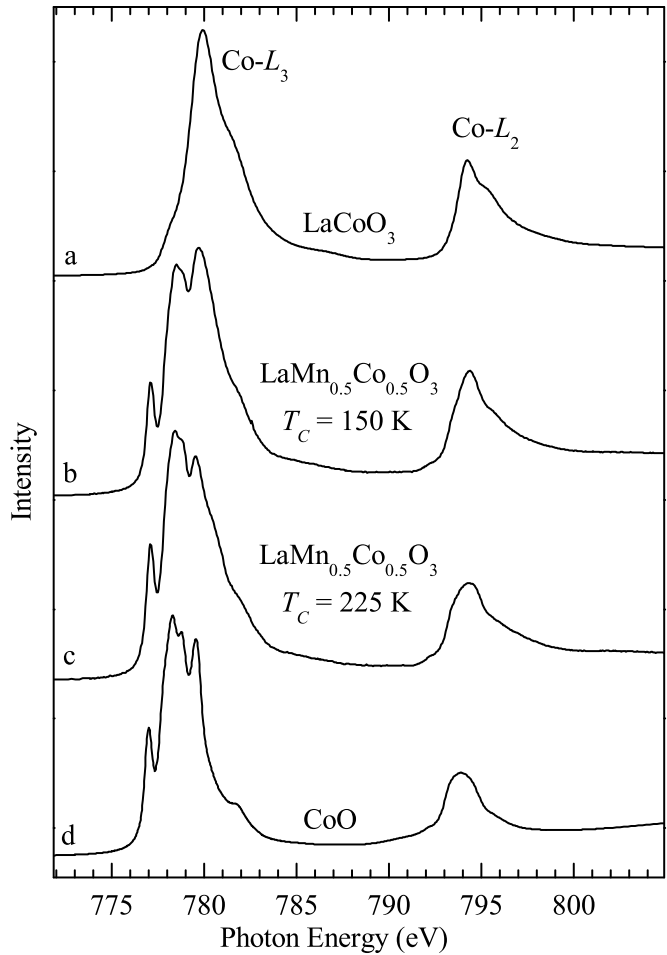}
    \caption{Co-$L_{2,3}$ XAS spectra of (a) LaCoO$_3$
    as a Co$^{3+}$ reference, of the
    LaMn$_{0.5}$Co$_{0.5}$O$_{3}$ samples with
    (b) $T_C = 150$~K and (c) $T_C = 225$~K,
    and (d) of CoO as a Co$^{2+}$ reference.}
    \label{fig:CoL23}
\end{figure}

\begin{figure}
    \includegraphics[width=0.5\textwidth]{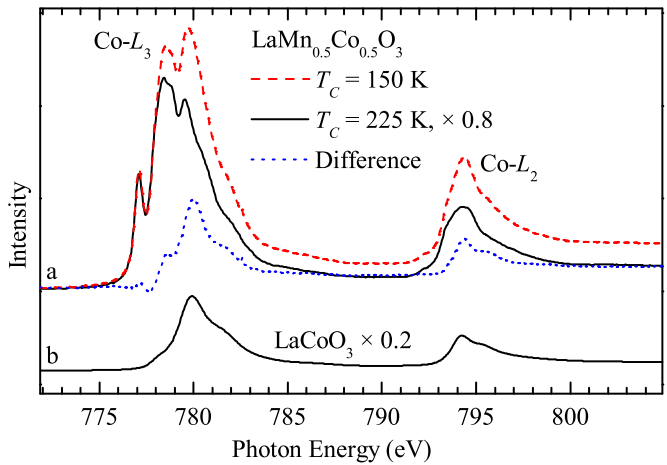}
    \caption{(Color online) Co-$L_{2,3}$ XAS spectra of
    (a) the LaMn$_{0.5}$Co$_{0.5}$O$_{3}$ samples
     with $T_C = 225$~K (solid black curve) and
    $T_C = 150$~K (dashed red curve), their difference
    (dotted blue curve), and (b) of
    LaCoO$_3$ as Co$^{3+}$ reference.}
    \label{fig:CoL23Comp}
\end{figure}

Figure~\ref{fig:CoL23} shows the Co-$L_{2,3}$ XAS spectra of
LaMn$_{0.5}$Co$_{0.5}$O$_{3}$ for both the high-$T_C$ [curve (c)]
and the low-$T_C$ phase [curve (b)]. The spectra were taken at room
temperature. For comparison, the spectrum of LaCoO$_3$ in the
low-temperature nonmagnetic state [curve (a)] is included as a
low-spin trivalent Co reference and also of CoO (curve d) as a
divalent Co reference. The spectra are dominated by the Co $2p$
core-hole spin-orbit coupling which splits the spectrum roughly
in two parts, namely the $L_{3}$ ($h\nu \approx 777$--780 eV) and
$L_{2}$ ($h\nu \approx 793$--796 eV) white lines regions. The
line shape of the spectrum depends strongly on the multiplet
structure given by the Co $3d$--$3d$ and $2p$--$3d$ Coulomb and
exchange interactions, as well as by the local crystal fields and
the hybridization with the O $2p$ ligands.

Important is that XAS spectra are highly sensitive to the valence
state: an increase of the valence state of the metal ion by one
causes a shift of the XAS $L_{2,3}$ spectra by one or more eV
toward higher energies.\cite{Chen90,Mitra2003} This shift is due
to a final state effect in the x-ray absorption process. The
energy difference between a $3d^n$ ($3d^7$ for Co$^{2+}$) and a
$3d^{n-1}$ ($3d^6$ for Co$^{3+}$) configuration is $\Delta
E=E(2p^63d^{n-1}{\to}2p^53d^{n})-E(2p^63d^n{\to}2p^53d^{n+1})\approx
U_{\underline pd}-U_{dd}\approx 1$--2~eV, where $U_{dd}$ is the
Coulomb repulsion energy between two $3d$ electrons and
$U_{\underline pd}$ the one between a $3d $ electron and the $2p$
core hole. In Fig.~\ref{fig:CoL23} we see a shift of the ``center
of gravity'' of the $L_3$ white line to higher photon energies by
approximately 1.5 eV in going from CoO to LaCoO$_3$. The energy
position and the spectral shape of the high-$T_C$ phase of
LaMn$_{0.5}$Co$_{0.5}$O$_{3}$ is very similar to that of CoO, indicating
an essentially divalent state of the Co ions.

While the spectral features of the low-$T_C$ phase of
LaMn$_{0.5}$Co$_{0.5}$O$_{3}$ are also very similar to those of
CoO and the high-$T_C$ phase as far as the low-energy side of the
$L_3$ white line is concerned, this is no longer true for the
high-energy side. The spectral weight at about 780 eV is increased when
one compares the high-$T_C$ with the low-$T_C$ phase, and this
increase is revealed more clearly by curves (a) of
Fig.~\ref{fig:CoL23Comp}. It is natural to associate this increase
with the presence of Co$^{3+}$ species since the LaCoO$_3$
spectrum has its main peak also at 780 eV. In order to verify
this in a more quantitative manner, we rescaled the spectrum of
the high-$T_C$ phase with respect to that of the low-$T_C$ phase
and calculate their difference. We find that a rescaling factor
of about 0.8 results in a difference spectrum (dotted blue curve
of Fig.~\ref{fig:CoL23Comp}) which resembles very much the
spectrum of LaCoO$_3$. This in turn may be taken as an indication
that the low-$T_C$ phase has about 20\% of its Co ions in the
low-spin trivalent state. This result contradicts the reports in
Refs.~\onlinecite{Joy2000a} and \onlinecite{Joly2001b} which
suggested that it was the high-$T_C$ sample which contained
trivalent Co ions.
The different result coming from the x-ray photoemission (XPS)
study\cite{Joly2001b} could be due to the following reason:
Unlike XAS in which the multiplet structure of the Co-$L_{2,3}$ spectra
is very characteristic for the Co valence, the XPS yields rather broad
and featureless Co $2p$ core-level spectra with very little distinction
between Co$^{2+}$ and Co$^{3+}$. To use XPS core-level shifts to
determine the valence state of insulating materials is also not so
straight forward due to the fact that the chemical potential with respect
to the valence or conduction band edges is not
well defined.
 The present finding of the presence of low-spin
Co$^{3+}$ species naturally explains why the low-$T_C$ sample has
less than the optimal $T_C$: the nonmagnetic ions suppress
strongly the spin-spin coupling between neighboring metal ions.

\begin{figure}
    \includegraphics[width=0.5\textwidth]{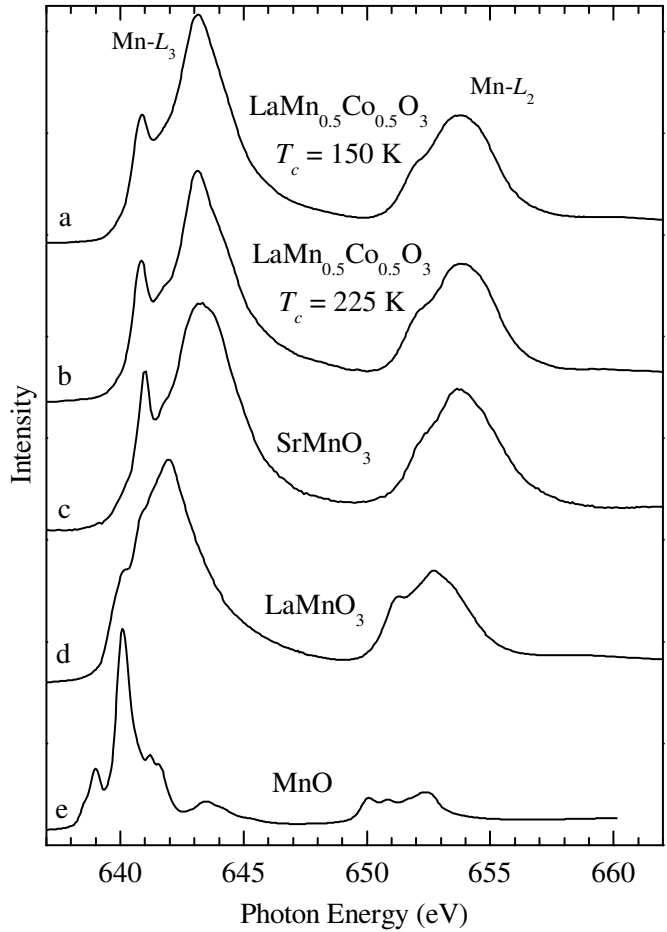}
    \caption{Mn-$L_{2,3}$ XAS spectra of the LaMn$_{0.5}$Co$_{0.5}$O$_{3}$ sample with
    (a) $T_C=150$~K and (b) $T_C=225$~K  together with
    (c) SrMnO$_3$ (Mn$^{4+}$, taken from Ref. \onlinecite{Sahu2002}),
    (d) LaMnO$_3$ (Mn$^{3+}$) and (e) MnO (Mn$^{2+}$) for comparison.}
    \label{fig:MnL23}
\end{figure}

\begin{figure}
    \includegraphics[width=0.5\textwidth]{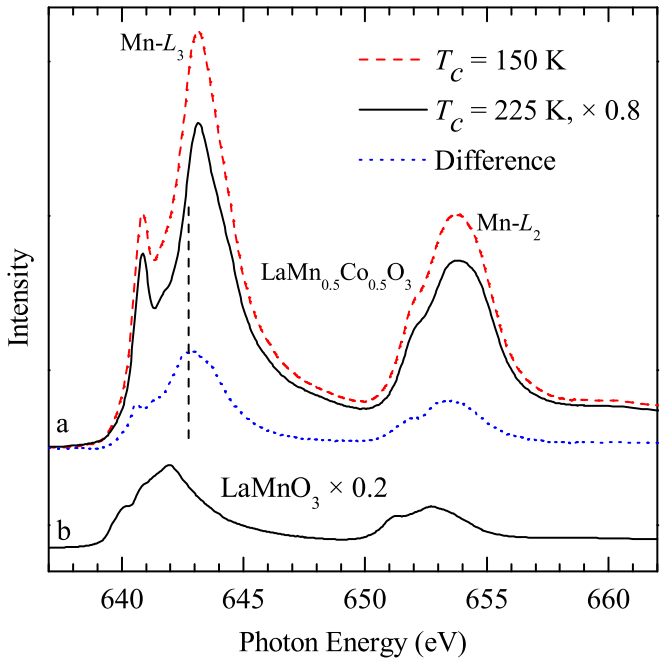}
    \caption{(Color online) Mn-$L_{2,3}$ XAS spectra of (a) the two LaMn$_{0.5}$Co$_{0.5}$O$_{3}$ samples with
    $T_C=150$~K (dashed red curve), $T_C=225$~K (black solid curve) and their difference (dotted blue curve), and
    (b) LaMnO$_3$ (Mn$^{3+}$) for comparison.}
    \label{fig:MnL23Comp}
\end{figure}

Figure~\ref{fig:MnL23} shows the room temperature Mn-$L_{2,3}$ XAS
spectra of the low-$T_C$ LaMn$_{0.5}$Co$_{0.5}$O$_{3}$ [curve (a)]
and the high-$T_C$ LaMn$_{0.5}$Co$_{0.5}$O$_{3}$ [curve (b)]
together with LaMnO$_3$ as a trivalent Mn reference [curve (c)] and
MnO as a divalent Mn reference [curve (d)]. Again we see a gradual
shift of the center of gravity of the $L_3$ white line to
higher energies from MnO to LaMnO$_3$ and further to SrMnO$_3$,
reflecting the increase of the Mn valence state from
$2+$ via $3+$ to $4+$. The Mn-$L_{2,3}$ spectrum of the
high-$T_C$ LaMn$_{0.5}$Co$_{0.5}$O$_{3}$ samples is similar to
that of SrMnO$_3$ and LaMn$_{0.5}$Ni$_{0.5}$O$_{3}$,\cite{Sanchez2002} in which
a Ni$^{2+}$/Mn$^{4+}$ valence state was found. The Mn-$L_{2,3}$
XAS spectrum thus reveals an essentially Mn$^{4+}$ state in the
high-$T_C$ LaMn$_{0.5}$Co$_{0.5}$O$_{3}$, consistent with the
observation of the Co$^{2+}$ valence in the Co-$L_{2,3}$ XAS
spectra above, i.e. fulfilling the charge balance requirement.

To investigate whether the presence of Co$^{3+}$ species in the
low-$T_C$ LaMn$_{0.5}$Co$_{0.5}$O$_{3}$ is also accompanied by
the occurrence of Mn$^{3+}$ ions as charge compensation, we have
carried out a similar analysis as for the Co spectra.
Figure~\ref{fig:MnL23Comp} shows the low-$T_C$ spectrum (red dashed
curve) and the high-$T_C$ one (black solid curve) rescaled to
80\% of low $T_C$. Their difference spectrum is shown as the
dotted blue curve. We find that the line shape resembles very much
that of the high-$T_C$ sample itself, suggesting that most of the
Mn in the low-$T_C$ sample are also tetravalent. This in turn
would imply that the low-$T_C$ sample has to have excess of oxygen to
account for the presence of the Co$^{3+}$ species. Nevertheless,
a closer look reveals that the energy position of the difference
spectrum lies between that of the Mn$^{4+}$ and the Mn$^{3+}$
spectra, and that the valley at 641--642 eV, at which energy a
typical Mn$^{3+}$ system like LaMnO$_3$ has its maximum, is not so
deep. This suggests that in the low-$T_C$ sample, there are also
some Mn$^{3+}$ ions or strongly hybridized Mn$^{3+}$ and Mn$^{4+}$
ions. Such a charge compensation for the Co$^{3+}$ could indicate
that the ordering of the Mn and Co ions is less than perfect, so
that the dislocated Co ions in the Mn$^{4+}$ positions would have smaller
metal-oxygen distances, leading to the stabilization of the
low-spin trivalent state of the Co.

\begin{figure}
    \includegraphics[width=0.5\textwidth]{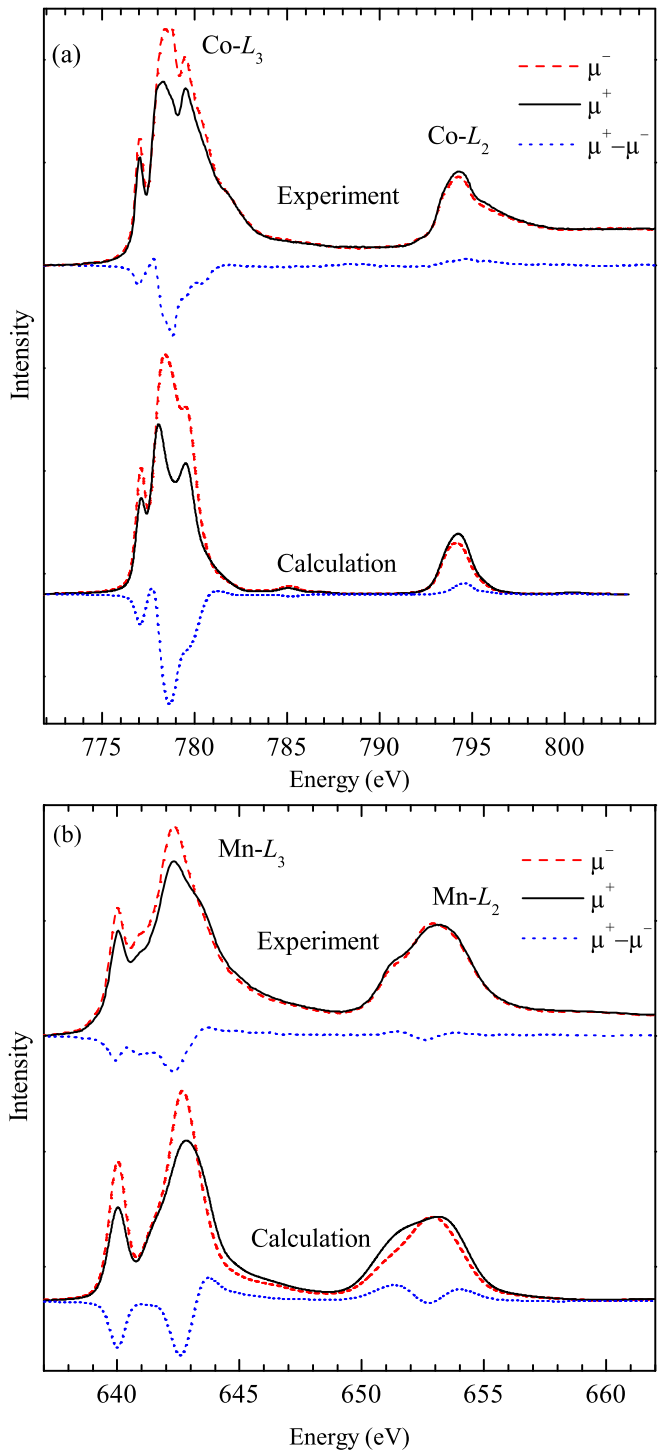}
    \caption{(Color online) Co-$L_{2,3}$ $(a)$ and Mn-$L_{2,3}$ $(b)$ spectra of LaMn$_{0.5}$Co$_{0.5}$O$_{3}$
     taken with circularly polarized x-rays at 135 K. The photon spin was aligned parallel ($\mu^+$, black solid)
      and antiparallel ($\mu^-$, red dashed) to the 1 T magnetic field, respectively;
      the difference spectra is shown in dotted blue.
    Top: measured spectra. Bottom: simulated spectra.}
    \label{fig:CoMnXMCD}
\end{figure}

Having established the valences of the Co and Mn ions, we now
focus our attention on their magnetic properties. In the top
panel (a) of Fig.~\ref{fig:CoMnXMCD}, we present the XMCD spectra
at the Co-$L_{2,3}$ edges of the high-$T_C$
LaMn$_{0.5}$Co$_{0.5}$O$_{3}$ taken at 135~K. The spectra
$\mu^{+}$ (black solid curve) and $\mu^{-}$ (red dashed curve)
stand, respectively, for parallel and antiparallel alignments
between the photon spin and the magnetic field. One can clearly
observe large differences between the two spectra with the
different alignments. The difference spectrum,
$\Delta\mu=\mu^{+}-\mu^{-}$, i.e. the XMCD spectrum, is also
shown (blue dotted curve). In the bottom panel (b) of
Fig.~\ref{fig:CoMnXMCD} we show the XMCD spectra at the
Mn-$L_{2,3}$ edges. Also here we can observe a large XMCD signal.
It is important to note that the XMCD is largely negative at both the Co and
the Mn $L_3$ edges, indicating that the Co$^{2+}$ and Mn$^{4+}$
ions are aligned ferromagnetically.

Very interesting about the XMCD at the Co-$L_{2,3}$ edges is that
it is almost zero at the $L_2$ while it is largely negative at
the $L_3$. This is a direct indication that the orbital
contribution ($L_z$, $m_{\rm orb}$) to the Co magnetic moment must
be large. In making this statement, we effectively used the XMCD
sum rule derived by Thole \textit{et al.},\cite{Thole1992} in
which the ratio between the energy-integrated XMCD signal and the
energy-integrated isotropic spectra gives a direct value for
$L_z$. Nevertheless, for a quantitative analysis it is preferred
to extract experimentally the $L_z$/$S_z$ ratio by making use of
an approximate XMCD sum rule developed by Carra \textit{et al.}
\cite{Carra1993} for the spin contribution (2$S_z$, $m_{\rm
spin}$) to the magnetic moment. This is more reliable than
extracting the individual values for $L_z$ and $S_z$ since one no
longer needs to make corrections for an incomplete magnetization,
due to, for example, possible strong magnetocrystalline
anisotropy in a polycrystalline material. The sum rules of Thole
\textit{et al.}\cite{Thole1992} and Carra \textit{et al.}
\cite{Carra1993} give for the $m_{\rm orb}$/$m_{\rm spin}$ or
$L_z$/2$S_z$,

\begin{align}\label{sumruleSzLz}
\frac{m_{\rm orb}}{m_{\rm spin}} &= \frac{L_z}{2S_z+7T_z} \nonumber\\
  &= \frac23 \frac{\int_{L_3} \Delta\mu(E)\,{\rm d}E + \int_{L_2} \Delta\mu(E)\,{\rm d}E}
               {\int_{L_3} \Delta\mu(E)\,{\rm d}E - 2 \int_{L_2} \Delta\mu(E)\,{\rm d}E},
\end{align}

\noindent where $T_z$ denotes the magnetic dipole moment. This
$T_z$ for ions in octahedral symmetry is a small number and
negligible compared to $S_z$.\cite{Teramura1996,note}\nocite{Crocombette1998}
Using this equation, we extract $m_{\rm orb}/m_{\rm spin} = 0.47$
out of our Co-$L_{2,3}$ XMCD spectrum.
This is a large value and is in fact
close to the value of 0.57 for CoO,\cite{Ghiringhelli2002} a
compound well known for the important role of the spin-orbit
interaction for its magnetic and structural
properties.\cite{Shull1951,Li1955,Roth1958,Laar1965,Khan1970,
Rechtin1970,Jauch2001,Kanamori1957,Nagayima1958,Shishidou1998}
The unquenched orbital moment is closely
related to the open $t_{2g}$ shell of the $3d^7$
configuration.\cite{Ballhausen,Csiszar2005}

Applying the sum rules for the Mn-$L_{2,3}$ XMCD spectra, we
obtain $m_{\rm orb}/m_{\rm spin} = 0.09$. This means that the
orbital moment for the Mn$^{4+}$ ions is nearly quenched.
Indeed, for the $3d^3$ configuration in the Mn$^{4+}$ compounds,
the majority $t_{2g}$ shell is fully occupied and thus a
practically quenched orbital moment is to be expected.

To critically check our findings concerning the local electronic
structure of the Co and Mn ions, we will explicitly simulate the
experimental XMCD spectra using the configuration interaction
cluster model.\cite{Tanaka94,deGroot94,Thole97} The method uses a
CoO$_6$ and MnO$_6$ cluster, respectively, which includes the
full atomic multiplet theory and the local effects of the solid.
It accounts for the intra-atomic $3d$--$3d$ and $2p$--$3d$ Coulomb
interactions, the atomic $2p$ and $3d$ spin-orbit couplings, the
oxygen $2p$--$3d$ hybridization, and local crystal field
parameters. Parameters for the multipole part of the Coulomb
interactions were given by the Hartree-Fock values,\cite{Tanaka94}
while the monopole parts ($U_{dd}$, $U_{pd}$) as well as the
oxygen $2p$--$3d$ charge transfer energies were determined from
photoemission experiments on typical Co$^{2+}$ and Mn$^{4+}$
compounds.\cite{Bocquet96} The one-electron parameters such as
the oxygen $2p$--$3d$ and oxygen $2p$--oxygen $2p$ transfer
integrals were extracted from band-structure calculations\cite{Wu07}
within the local-density approximation (LDA)
using the low-temperature crystal
structure of the high-$T_C$ phase.\cite{Troyanchuk2004} The
simulations have been carried out using the \textsc{XTLS 8.3}
program\cite{Tanaka94} with the parameters given in Ref.
\onlinecite{parameters}.

Important for the local electronic structure of the Co$^{2+}$ ion
is its local $t_{2g}$ crystal field scheme. This together with
the spin-orbit interaction determines to a large extent its
magnetic properties. To extract the crystal field parameters
needed as input for the cluster model, we have performed
constrained LDA+U calculations\cite{Wu07}
without the spin-orbit interaction. 
We find that the $zx+xy$ orbital lies lowest, while
the $yz$ is located 22 meV and the $zx-xy$ 27 meV higher. Here, we
made use of local coordinates in which the $z$ direction is along
the long Co--O bond (2.078 \AA), the $y$ along the second-longest
bond (2.026 \AA), and the $x$ along the short bond (1.997 \AA).
The cluster model finds the easy axis of the magnetization to lie
in the $yz$ direction with a single-ion anisotropy energy of
about 0.5--1.5 meV, i.e., larger than can be achieved by the
applied magnetic field. Since we are dealing with a
polycrystalline sample, the sum of spectra taken with the light
coming from all directions has to be calculated; we approximated
this by summing two calculated spectra: one for light with the
Poynting vector along the $yz$ axis and one with the Poynting
vector perpendicular to this. The exchange field direction is
kept along the $yz$ in both cases.

The results of the cluster model calculations are included in
Fig.~\ref{fig:CoMnXMCD}, in the top panel (a) for the Co
$L_{2,3}$ edges and the bottom panel (b) for the Mn. One can see
that the line shapes of the experimental Co and Mn spectra are
well explained by the simulations: all the characteristic features
are reproduced. We would like to remark that the experimental
XMCD spectra (blue dotted curves) are in general about 30\%
smaller than the simulated XMCD spectra (blue dotted curves).
This is due to the fact that the experimental spectra were \textit{not}
corrected for the incomplete degree of circular polarization
($\approx\!80\%$) of the beamline, nor for the fact that magnetic
field makes an angle of $30^\circ$ with respect to the Poynting
vector of the light, nor for the reduction of the magnetization at 135~K
at which the sample was measured -- compared to the calculation which were
done at 0~K. From these simulations we thus can safely
conclude that our interpretation for the Co and Mn valences and
magnetic moments is sound.

\begin{figure}
    \includegraphics[width=0.5\textwidth]{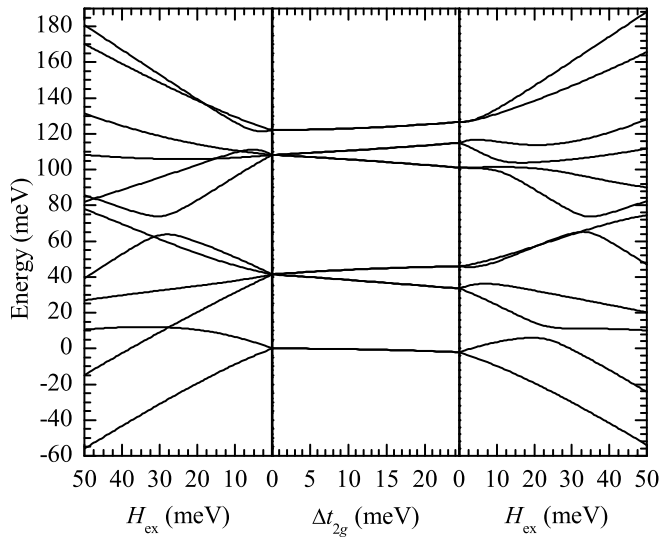}
    \caption{Energy level diagram of the Co$^{2+}$ ion
             (left panel) in a cubic field depending on the
             strength of the exchange field $H_{\rm ex}$,
             (middle panel) the effect of lowering the symmetry
             [$\Delta e_g=4\Delta t_{2g}=4(E_{zx-xy,yz}-E_{zx+xy})$],
             and (right panel) the low-symmetry energy splitting depending on
             $H_{\rm ex}$.}
    \label{fig:energy_level}
\end{figure}

Our finding of a large orbital contribution to the Co magnetic
moment has important implications for the interpretation of the
magnetic susceptibility data. In most of the studies published so
far, one tried to extract magnetic quantum numbers from the
magnetic susceptibility data using the Curie or Curie-Weiss law by
finding a temperature region in which the inverse of the magnetic
susceptibility is linear with temperature. One usually takes the
high temperature region. We will show below that this standard
procedure will {\em not} provide the magnetic quantum numbers
relevant for the ground state of this material.

The fact that the $3d$ spin-orbit interaction in this Co material
is ``active'' has as a consequence that the energy difference
between the ground state and the first excited state will be of
the order of the spin-orbit splitting $\zeta$, which is about 66
meV for the Co$^{2+}$ ion. We have illustrated this in
Fig.~\ref{fig:energy_level} which shows the energy level diagram
of the Co$^{2+}$ ion, both in cubic symmetry (left panel) and in
the low-temperature and ferromagnetic state of the
LaMn$_{0.5}$Co$_{0.5}$O$_{3}$ system (right panel) where we have
used the crystal field scheme as described above.

\begin{figure}
    \includegraphics[width=0.5\textwidth]{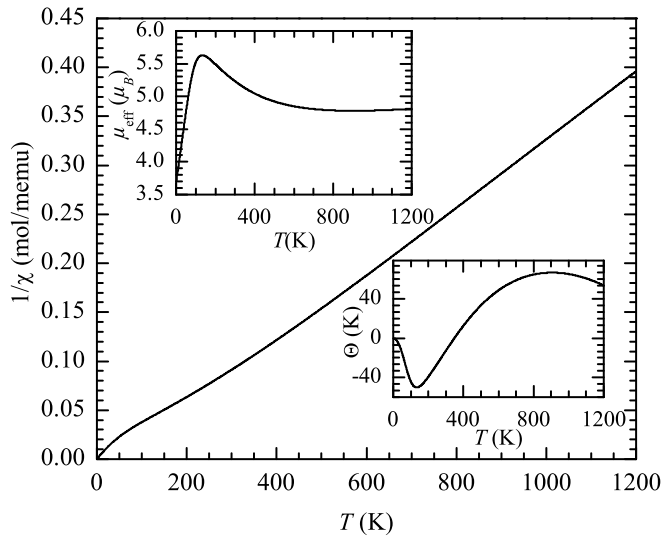}
    \caption{Calculated inverse susceptibility for a single Co$^{2+}$
             ion in a cubic crystal field; (top inset) the (apparent) effective
             moment $\mu_{\textrm{eff}}$ and (bottom inset) the
             (apparent) Weiss temperature $\Theta$ as defined in the text.}
    \label{fig:chi_mu_theta}
\end{figure}

To demonstrate the consequences of the presence of such a set of
low lying excited states, we calculated the magnetic
susceptibility $\chi$ of the Co$^{2+}$ in cubic symmetry for an
applied magnetic field of 0.01 T and without an exchange
field. The results are presented in Fig.~\ref{fig:chi_mu_theta}
where we depict also the (apparent) effective magnetic moment
$\mu_{\textrm{eff}}$ [$\mu_{\textrm{eff}}^2$ is defined here as
$3k_B$ divided by the temperature derivative of $1/\chi(T)$] and the
(apparent) Weiss temperature $\Theta$ [$\Theta$ is defined here
as the intercept of the tangent to the $1/\chi(T)$ curve with the
abscissa]. One can clearly observe that $1/\chi(T)$ is not linear
with temperature for temperatures between $T_C = 225$ K and
roughly 800 K. Only for temperatures higher than 800 K, one can
find a Curie-Weiss-like behavior, but then the (apparent) Weiss
temperature has nothing to do with magnetic correlations since
they were not included in this single ion calculations. Instead,
the (apparent) Weiss temperature merely reflects the fact that
the first excited states are thermally populated. This means in
turn that one cannot directly extract the relevant
\textit{ground} state quantum numbers from the high temperature
region.

In principle, one could hope to find a Curie-Weiss behavior by
focusing on the very low temperature region only, e.g., below 50
K, but there one has to take into account that there is a very
large van Vleck contribution to the magnetic susceptibility due
to the fact that the first excited states are lying very close,
i.e., in the range of the spin-orbit splitting. The extrapolation
to $T=0$ K would then give the real value for $\mu_{\textrm{eff}}$
of the ground state. In the case of LaMn$_{0.5}$Co$_{0.5}$O$_3$,
however, the presence of ferromagnetism, which already sets in at
225 K, will completely dominate the magnetic susceptibility and
thus hinder the determination of $\mu_{\textrm{eff}}$ of the
ground state using this procedure. Obviously, one can determine in
principle the magnetic moments in a ferromagnet from the
saturation magnetization, but apparently this is the issue for
LaMn$_{0.5}$Co$_{0.5}$O$_3$ where one is debating about the
importance of Mn/Co disorder and its relationship to reduced
magnetizations and less than optimal Curie temperatures.

Another often used ``magnetic'' technique to determine the moments
in this ferromagnetic material is neutron scattering. Troyanchuk
\textit{et al.} found a mean value of $2.5\mu_B$ per formula
unit (LaCo$_{0.5}$Mn$_{0.5}$O$_3$).\cite{Troyanchuk2004} The
authors claimed that this is in good agreement with the
Co$^{2+}$--Mn$^{4+}$ scenario. Indeed, assuming spin-only moments
as is generally done (but which is not correct as shown above),
one would already expect $3\mu_B$ for a Co$^{2+}$ ion and
$3\mu_B$ for a Mn$^{4+}$ ion, totaling to $6\mu_B$, i.e.
$3\mu_B$/f.u., which is somewhat larger than the
experimental finding and which can be understood consistently if
one assumes that the Co--Mn ordering in their sample is not
perfect. It is important to note that the low-spin Co$^{3+}$--Mn$^{3+}$
scenario can be ruled out since this yields only $2\mu_B$/f.u.,
i.e., too low to explain the experiment.
Nevertheless, a Co$^{3+}$--Mn$^{3+}$ scenario in which the
Co$^{3+}$ ion is in the intermediate ($S=1$) or high spin state
($S=2$) cannot be excluded on the basis of the moments measured
by the neutrons alone.\cite{Hu04,Haverkort2006}

Our cluster model calculations based on the XAS and XMCD spectra
reveal that the Co$^{2+}$ ion has $m_{\rm spin} = 2.12\mu_B$ and
$m_{\rm orb} = 0.99\mu_B$ and that the Mn$^{4+}$ has $m_{\rm
spin} = 2.84\mu_B$ and $m_{\rm orb} = 0.02\mu_B$, totaling to
$2.99\mu_B$/f.u. This is
not inconsistent with the magnetization results of Asai
\textit{et al.},\cite{Asai1998} if we make an extrapolation to
higher magnetic fields as to estimate the
saturated total moment. Our result is
larger than the neutron results, but also
not inconsistent if one is willing to accept
that there is an appreciable amount of Co--Mn disorder in the
neutron sample. Crucial is that our XAS and XMCD spectra rule out
\textit{all} the Co$^{3+}$--Mn$^{3+}$ scenarios: (1) our Co
$L_{2,3}$ spectra give a positive match with those of
Co$^{2+}$ compounds, while they do not fit those of low-spin
Co$^{3+}$ and high-spin Co$^{3+}$
compounds;\cite{Hu04,Haverkort2006} (2) our Mn
$L_{2,3}$ spectra are very similar to those of Mn$^{4+}$
compounds, and very dissimilar to those of Mn$^{3+}$.

To summarize, we have utilized an element-specific spectroscopic
technique, namely, soft-x-ray absorption and magnetic circular
dichroism spectroscopy, to unravel the local electronic structure
of LaMn$_{0.5}$Co$_{0.5}$O$_{3}$ system. We have firmly
established the high-spin Co$^{2+}$--Mn$^{4+}$ scenario. We have found a very large orbital contribution to
the Co magnetic moment, implying a nontrivial temperature
dependence for the magnetic susceptibility. We also have revealed
that samples with lower Curie temperatures contain low-spin
nonmagnetic Co$^{3+}$ ions.

We would like to thank Daniel Khomskii for critical reading the
paper and Lucie Hamdan for her skillful technical and
organizational assistance in preparing the experiments. The
research in Cologne is supported by the Deutsche
Forschungsgemeinschaft through SFB 608.

\end{document}